\documentclass{acm_proc_article-sp}

\usepackage{algorithmic}
\usepackage{footnote}
\usepackage{url}
\usepackage[caption=false,font=footnotesize]{subfig}

\begin{document}
\title{MDMP: Managed Data Message Passing}

%
%
%
%
%

\numberofauthors{2} 
%
\author{
%
%
\alignauthor Adrian Jackson\\
\affaddr{EPCC} \\
\affaddr{The University of Edinburgh}\\
\affaddr{Mayfield Road}\\
\affaddr{Edinburgh}\\
\affaddr{EH9 3JZ, UK}\\
\email{Adrian.Jackson@ed.ac.uk}  
\alignauthor P\"{a}r Strand \\
\affaddr{Plasma physics and fusion energy} \\
\affaddr{Earth and Space Sciences} \\
\affaddr{Chalmers University} \\
\affaddr{H\"{o}salsv\"{a}gen 9, 4th floor} \\
\affaddr{G\"{o}teborg, Sweden} \\
\email{par.strand@chalmers.se}
}
\maketitle
\begin{abstract}
MDMP is a new parallel programming approach that aims to provide users with an easy way to add parallelism to programs, optimise the message passing costs of traditional scientific simulation algorithms, and enable existing MPI-based parallel programs to be optimised and extended without requiring the whole code to be re-written from scratch. MDMP utilises a directives based approach to enable users to specify what communications should take place in the code, and then implements those communications for the user in an optimal manner using both the information provided by the user and data collected from instrumenting the code and gathering information on the data to be communicated. In this paper we present the basic concepts and functionality of MDMP and discuss the performance that can be achieved using our prototype implementation of MDMP on some simple benchmark cases.
\end{abstract}


\section{Introduction}
\label{sec:0}
There are numerous new programming languages, libraries, and techniques that have been developed over the past few years to either simplify the process of developing parallel programs or provide additional functionality 
that traditional parallel programming techniques (such as MPI\cite{mpi2} or OpenMP\cite{openmp}) do not provide.  These include programming extensions such as Co-Array FORTRAN\cite{coarray} (CAF), UPC\cite{upc}, and new 
languages such as Chapel\cite{chapel}, OpenMPD\cite{openmpd} or XMP\cite{xmp}

However, these approaches often have not has a focus of optimising the parallelisation overheads (the cost of the communications) associated with distributed memory parallelisation, and have limited appeal for the many 
scientific applications which are already parallelised with an existing parallelisation approach (primarily MPI) and have very large code bases which would be prohibitively expensive to re-implement in a new parallel 
programming language.  

The challenge of optimising parallel communications is becoming increasingly important as we approach Exascale-type high performance computers (HPC), where it is looking increasingly likely the that ratio between the 
computational power of a node of the computer and the relative performance of the network is going to make communications increasingly expensive when compared to the cost of calculations.  Furthermore, the rise of 
multi-core and many-core computing on the desktop, and the related drop in single core performance, means that many more developers are going to need to exploit parallel programming to utilise the computational resources 
they have access to where they could have relied on increases in serial performance of the hardware they were using to maintain program performance in the past.  Therefore, we have devised a 
new parallel programming approach, called Managed Data Message Passing (MDMP)\cite{mdmp}, which is based on the MPI library but provides a new method for parallelising programs. 

MDMP follows the directives based approach, favoured by OpenMP and other parallel programming techniques, which are translated into MDMP library function calls or code snippets, which in turn utilise communication library 
calls (such as MPI) to provide the actual parallel communication functionality.  Using a directives based approach enables us to reduce the complexity, and therefore development code, of writing parallel programs, 
especially for the novice HPC programmer.  However, the novel aspect of MDMP is that it allows users to specify the communication patterns required in the program but devolves the responsibility 
for scheduling and carrying out the communications to the MDMP functionality.  MDMP instruments data accesses for data being communicated to optimise when communications happen and therefore better overlap 
communication and computation than is easily possible with traditional MPI programming.  

Furthermore, by taking the directive approach MDMP can be incrementally added to a program that is already parallelised with MPI, replacing or extending parts of the existing MPI parallelisation without requiring any 
changes to the rest of the code.  Users can start by replacing one part of the current communication in the code, evaluate the performance impacts, and replace further communications as required.

In this paper we outline work others have undertaken in creating new parallel programming techniques, and optimisation communications.  We describe the basic issues we are looking to tackle with MDMP, and go on to describe 
the basic feature and functionality of MDMP, outlining the performance benefits and costs of such an approach, and highlighting the scenarios where MDMP can provide reduced communication 
costs for the types of communication patterns seen in some scientific simulation codes using our prototype implementation of MDMP (which implements MDMP as library calls rather than directives).  

\section{Related Work}

Recent evaluation of the common programming languages used in large scale parallel simulation code has found the majority are still implemented using MPI, with a minority also including a hybrid parallelisation 
through the addition of OpenMP (or in a small number of cases SHMEM) alongside the MPI functionality\cite{pracesurvey}.  This highlights to us the key requirement for any new parallel programming language or technique of 
being easily integrated with existing MPI-based parallel codes.

There are a wide range of studies evaluating the performance of a range of different parallel languages, including Partitioned Global Address Space (PGAS) languages, on different application 
domains and hardware\cite{Wen:2007:AMR:1362622.1362676, Jin:2009:PSU:1809961.1809973, Shan:2010:PMP:1938482.1938487,Blagojevic:2010:HPR:2020373.2020376}.  These show that there are many approaches 
that can provide performance improvements for parallel programs, compared to standard parallelisation techniques on a given architecture or set of architectures.   Existing parallel programming 
languages or models for distributed memory system provide various features to 
describe parallel programs and to execute them efficiently.  For instance, XMP provides features similar to 
both CAF and HPF\cite{Muller:1995:EHP:224538.224552}, allowing users to use either global or local view 
programming models, and providing easy to program functionality through the use of compiler directives for 
parallel functionality.  Likewise, OpenMPD provided easy to program directives based parallelisation for 
message passing functionality, extending an OpenMP like approach to a distributed memory supercomputer.

However, both of these approaches generally require the re-writing of existing codes, or parts of existing codes, into a new languages, 
which we argue is prohibitively expensive for most existing computational simulation applications and therefore has limited the take-up 
of these different parallel programming languages or techniques by end user applications.  Furthermore, both only target parts of the 
problem we are aiming to tackle, namely improving programmability and optimising performance.  XMP and OpenMPD both aim to make parallel 
programming simpler, but have not direct features for optimising communications in the program (although they can enable users to implement 
different communication methods and therefore choose the most efficient method for themselves).  PGAS languages, and other new languages, 
may provide lower cost communications or new models of communications to enable different algorithms to be used for a given problem, 
or may provide simpler programming model, but none seems to offer both as a solution for parallel programming.  Also, crucially, they 
do not expect to work with existing MPI programs, negating the proposed benefits for the largest part of current HPC usage.

There has also been significant work undertaken looking at optimising communications in MPI programs.  A number of authors have looked at 
compiler based optimisations to provide automatic overlapping of communications and computation in existing parallel 
programs\cite{staticoverlap,Fishgold:2006:AAI:1898699.1898816,Iancu:2008:PPO:1375527.1375567}.  These approaches have shown that 
performance improvements can be obtained, generally evaluated against kernel benchmarks such as the NAS parallel benchmarks, by transforming user specified blocking 
communication code to non-blocking communication functionality, and using static compiler analysis to determine where the communications 
can be started and finished.  Furthermore, other authors have looked at communication patterns or models in MPI based parallel programs 
and suggested code transformations that could be undertaken to improve communication and computation overlap\cite{Danalis05transformationsto}. 
 However, these approaches are what we would class as {\it coarse-grained} communication optimisation.  They use only static compiler analysis 
to identify the communication patterns, and identify the outer bounds of where communications can occur to try and start and finish bulk 
non-blocking operations in the optimal places.  They do not address the fundamental separation of communication and computation into different 
phases that such codes generally employ.  Our work, outlined in this paper, is looking at {\it fine-grained} communication optimisations, where 
individual communication calls are {\it intermingled} with computation to truly mix communication and computation.

The has also been work on both offline and runtime identification and optimisation of MPI communications, primarily for collective communication\cite{Hoefler:2012:RDO:2370816.2370856,detectcollectmpidongarra,Faraj:2006:SST:1183401.1183431}, or other auto-tuning techniques such optimising MPI library variables\cite{Pellegrini:2010:ATM:1787275.1787310} or individual library routines\cite{Li:2009:NAG:1561015.1560858}.  All these approaches have informed the way we have constructed MDMP.

We belief that the work we have undertaken is unique as it brings together attempts to provide simple message passing programming which fine-grained 
communication optimisation along with the potential for runtime auto-tuning of communication patterns into a single parallel programming tool.

\section{Communication Optimisation}

\begin{figure}[t]
\centering
\epsfig{file=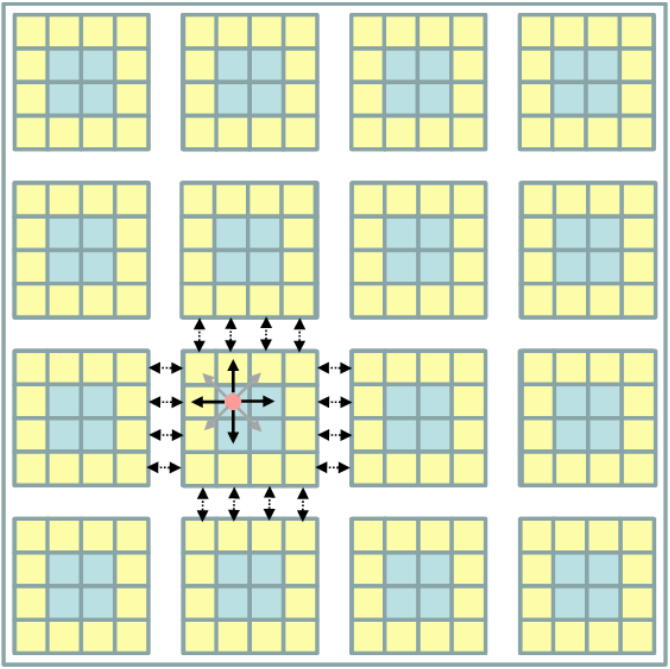,width=0.5\textwidth}
\label{fig:commpattern}
\caption{Representation of a common communication pattern for parallel simulations}
\end{figure}

Whilst there is a very wide range of communication and computational patterns in parallel programs, a large 
proportion of common parallel applications use regular domain decomposition techniques coupled with {\it halo} 
communications to exploit parallel resources.  As shown in Figure\ref{fig:commpattern}, which is a 
representation of a Jacobi-style stencil based simulation method, many simulations undertake a set of 
calculations that iterate over a n-dimensional array, with a set of communications to neighbouring 
processes every iteration of the simulation.

The core computational kernel of a simple Jacobi style simulation, as illustrated in the previous paragraph, 
can be implemented as shown in Figure \ref{code:original} (undertaking a 2d simulation).
\begin{figure}
\begin{verbatim}
for (iter=1;iter<=maxiter; iter++){      
   MPI_Irecv(&old[0][1], NP, MPI_FLOAT, prev, 1, 
      MPI_COMM_WORLD, &requests[0]);
   MPI_Irecv(&old[MP+1][1], NP, MPI_FLOAT, next, 2, 
      MPI_COMM_WORLD, &requests[1]);      
   MPI_Isend(&old[MP][1], NP, MPI_FLOAT, next, 1, 
      MPI_COMM_WORLD, &requests[2]);
   MPI_Isend(&old[1][1], NP, MPI_FLOAT, prev, 2, 
      MPI_COMM_WORLD, &requests[3]);      
   MPI_Waitall(4, requests, statuses);      
   for (i=1;i<MP+1;i++){
      for (j=1;j<NP+1;j++){
         new[i][j]=0.25*(old[i-1][j]+old[i+1][j]+
	    old[i][j-1]+old[i][j+1] - edge[i][j]);
      }
   }
   for (i=1;i<MP+1;i++){
      for (j=1;j<NP+1;j++){
         old[i][j]=new[i][j];
      }
   }
}
\end{verbatim}
\label{code:original}
\caption{MPI implementation of a simple 2d Jacobi style computation, implemented in C using MPI}
\end{figure}

It is evident from the above code that, whilst it has been optimised to use non-blocking communications, the 
communication and computation parts of the simulation are performed separately, with no opportunity to overlap 
communications and computations.  In practice this means that the application will only be using the 
communication network to send and receive data in short bursts, leaving it idle whilst computation is being 
performed.  

Many large scale HPC resources are used by a large number of running applications at any one time, which may 
help to ensure that the overall usage of the interconnect is high, even though individual applications often 
utilise it in a {\it bursty} manner. However, that still will not be true of the part of the network 
dedicated to the individual application, only to the load of the network overall.  Furthermore, when 
considering the very largest HPC resources in the world, and including the proposed Exascale resources, 
there are often only a handful of applications utilising the resource at any one time.  
Therefore, enabling applications to effectively utilise the network, especially the {\it spare} resources 
that the current separated communication and computation patterns engender, is likely to be beneficial 
to overall application performance and resource utilisation (provided that the cost of doing this is not significant).

It is possible to split the sends and receives in the previous example and place them around the computation rather than just before the computation, using the non-blocking 
functionality, to further ensure that more optimal communications are occurring.  However, this is still does not allow for overlapping communication and computation 
because the computational is still occurring in a single block, with communications outside this block.

For developers to ensure that communications and computations are truly mixed would require further code modifications, as shown in the code example in 
Figure \ref{code:optmpi} (which implements a strategy of sending data as soon as it has been computed).

\begin{figure}
\small
\begin{verbatim}
for (iter=1;iter<=maxiter; iter++){
    requestnum = 0;
    for (j=0;j<NP;j++){
        MPI_Irecv(&tempprev[j], 1, MPI_FLOAT, prev, 1, 
           MPI_COMM_WORLD, &requests[requestnum]);
        requestnum++;
        MPI_Irecv(&tempnext[j], 1, MPI_FLOAT, next, 2, 
           MPI_COMM_WORLD, &requests[requestnum]);
        requestnum++;
    }
    for (i=1;i<MP+1;i++){
        for (j=1;j<NP+1;j++){
            new[i][j]=0.25*(old[i-1][j]+old[i+1][j]+
               old[i][j-1]+old[i][j+1] - edge[i][j]);
            if(i == MP){
               MPI_Isend(&new[i][j], 1, MPI_FLOAT, next, 1,
                  MPI_COMM_WORLD, &requests[requestnum]);
               requestnum++;
            }else if(i == 1){
                MPI_Isend(&new[i][j], 1, MPI_FLOAT, prev, 2,
                   MPI_COMM_WORLD, &requests[requestnum]);
                requestnum++;
            }
        }
    }
    for (i=1;i<MP+1;i++){
        for (j=1;j<NP+1;j++){
            old[i][j]=new[i][j];
        }
    }
    MPI_Waitall(requestnum, requests, statuses);
    if(prev != MPI_PROC_NULL){
       for (j=1;j<NP+1;j++){
           old[0][j] = tempprev[j-1];
           old[MP+1][j] = tempnext[j-1];         
       }
    }
    if(next != MPI_PROC_NULL){
       for (j=1;j<NP+1;j++){
           old[MP+1][j] = tempnext[j-1];         
       }
    }
}
\end{verbatim}
\normalsize
\label{code:optmpi}
\caption{Intermingled MPI implementation of a simple 2d Jacobi style computation}
\end{figure}

Whilst the code implemented in Figure \ref{code:optmpi} will enable the mixing of communication and computation, ensuring that data is sent as 
soon as it is ready to be communicated and potentially ensuring better utilisation of the communication network, it has come at the cost of considerable code {\it mutilation}, 
requiring developers to undertake significant code optimisations.  As well as the damage to the readability and 
maintainability of the code that this causes, it also means that a code has been significantly changed for a 
potentially architecturally dependent optimisation, i.e. an optimisation that may be beneficial on one or more current HPC systems but may not be beneficial on other or future HPC systems.

We are proposing MDMP as a mechanism for implementing such optimisations without the requirement to significantly change users codes, or the need to tailor codes to a specific platform, as the MDMP functionality can 
implement communications in the most optimal form for the application and hardware currently being used.

\section{MDMP}

To re-iterate the challenges for MDMP that we have previously discussed, MDMP is designed to address the following issues:
\begin{itemize}
\item Work with existing MPI based codes
\item Provide framework for optimisation communications
\item Simplify parallel development
\end{itemize}

MDMP uses a directives based approach, relying on the compiler to implement the actual message passing 
functionality based on the users' instructions.  Compiler directives are used, primarily, to address the 
third point above, namely ease of use.  We provide functionality that can be easily enabled and disabled 
in an application, hides some of the complexities of current MPI programming (such as providing message tags, 
error variables, communicators, etc...) that often complicate development for new users of MPI, and 
also provides some flexibility to the user over the type and level of message optimisation used.

The MDMP directives are translated into code snippets and library calls by the MDMP-enabled compiler, 
either directly in the equivalent non-blocking MPI calls (which simply mimics the communication 
that would have been implemented directly by the user) or to further optimised MPI communications or 
other another communication library as appropriate on the particular hardware being used.  This enables 
MDMP to target different communication libraries transparently to the developer for a given HPC system. 
Also, crucially the ability to target MPI communications means that MDMP functionality can be added to 
existing MPI-parallelised programs, either as additional functionality or to replace existing MPI functionality, 
without requiring the program to be completely changed into a new programming language or utilise a 
new message-passing (or other) communication library.

However, simply using directives for programming message passing will not optimise the communication that are 
undertaken by a program.  Therefore, MDMP provides not only directives to specify the communications to be 
undertaken in the program but also directives to specify {\it communication regions}.  Communication regions 
define the areas of code where the data that is to be sent and received is worked on, and where communications 
occur, so that MDMP can, at runtime, examine the data access patterns and undertake communications at the 
optimal time to intermingle communications and computations and therefore better utilise the communication 
network.

The optimisation of communications is based on runtime functionality that monitors the reads and writes of 
data that has been specified as communication data (data that will be sent or received).  As any data monitoring 
entails some runtime overheads the communication region specifies the scope of the data monitoring to ensure 
it is only performed where required (i.e. where communications are occurring).  Any data that is specified 
by the users and being involved in send or receives is tracked so each read and writing in a communication 
region is recorded and the number of reads and writes that have occurred when the send or receive happens is 
evaluated.  This data, the number of reads and writes that have occurred for a particular piece of data when it 
comes to be sent or written over by a receive, can then be used in any subsequent iterations of the 
computation to launch the communication of that data once it is ready to be communicated.

Communications are triggered for any given piece of data as follows:
\begin{itemize}
\item last write occurs (sends)
\item last read and/or write occurs (receives)
\end{itemize}

Using this functionality we can implement a communication pattern that intermingles communication and 
computation for the example code shown in Figure \ref{code:original}, as shown in Figure \ref{code:mdmp}.

\begin{figure}
\begin{verbatim}
#pragma commregion
for (iter=1;iter<=maxiter; iter++){      
#pragma recv(old[0][0], NP, prev)
#pragma recv(old[MP+1][1], NP, next)
#pragma send(old[MP][1], NP, next)
#pragma send(old[1][1], NP, prev)
   for (i=1;i<MP+1;i++){
       for (j=1;j<NP+1;j++){
           new[i][j]=0.25*(old[i-1][j]+old[i+1][j]+
              old[i][j-1]+old[i][j+1] - edge[i][j]);
       }
   }
   for (i=1;i<MP+1;i++){
       for (j=1;j<NP+1;j++){
           old[i][j]=new[i][j];
       }
   }
}
#pragma commregionfinished	
\end{verbatim}
\label{code:mdmp}
\caption{MDMP implementation of a simple 2d Jacobi style computation with optimised communication}
\end{figure}

When compiled with an MDMP-enabled compiled, the code in Figure \ref{code:mdmp} will be processed by the 
compiler and non-blocking sends and receives inserted where the \verb|send| and \verb|recv| directives are placed. 
The compile then looks through the code associated with the communicating region (between \verb|commregion| and 
\verb|commregionfinished|) and replaces any variable reads or writes linked to those sends and receives by MDMP 
code which will perform the reads and writes and also record those reads and writes occurring.

Compiler based code analysis for data accesses will be straightforward for many applications, however we recognise that 
there will be a number of scenarios, such as when pointers are heavily used in C or FORTRAN, or possibly where pre-processing 
or function pointers or conditional function calls are used, where it will not be possible 
for the compiler to access where the data accesses for a particular \verb|send| or \verb|recv| occur.  In that 
situation MDMP will revert to simply inserting the basic MPI function calls required to undertake the specified communication 
and not perform the optimise message functionality.  Whilst this negates the possibility of optimising the communications, it 
will not add any overheads to the program compared to the standard MPI performance a developer would experience, and it does 
still leave scope for the MDMP functionality to target communication libraries other than MPI to enable optimisation for 
users would requiring them to modify their code, if such functionality is available.

Furthermore, whilst we are not investigating such functionality at the moment, the design of MDMP means that it can also 
undertake auto-tuning or other runtime activities to optimise communication performance for users beyond the intermingling 
communication optimisations we have already discussed.  For instance, MDMP could implement additional helper threads that 
enable progression of communications whilst the main program is undertaking calculations, albeit at the cost of utilising 
a computational core for that purpose.  It could also evaluate different communication optimisations at runtime to auto-tune 
the performance of the program whilst it is running.

A difference between the MPI functionality that a developer would add to a code like the one we have been considering and 
the functionality that MDMP implements is that where intermingling of communications is undertaken MDMP will be sending lots 
of single element (or small numbers of elements) messages between processes rather than a single message with all the data in it.
In general, MPI performs best when small numbers of large messages are used, rather 
than large numbers of small messages.  This is because in the case that large numbers of small message are sent the communication costs 
are dominated by the latency costs of each message, whereas using small numbers of large messages reduces the overall number of message 
latencies that are incurred.  We recognise the fact the the MDMP functionality may not be optimal in terms of the overall 
message costs associated with communications but we are assuming that this penalty will be negated by the benefits associated with 
more consistent use of the network and less concentrated nature of the communication and computation patterns.  However, this is something 
that is investigated in our performance analysis of MDMP, and as with other previously discussed potential problems with MDMP if it 
is impacting performance the optimised message passing can be disabled at compile time.

We also recognise that MDMP functionality does not come without a cost to the performance of the program.  MDMP is adding 
additional computational requirements above those specified in the user program, and also requires additional memory 
to store data associated with the communications (such as the counters that record the reads and writes to variables).  
The premise behind the optimised message passing functionality we are aiming for is that communications are much more 
expensive than computations for an application on a modern HPC machine, and this relationship is likely to get worse for 
future HPC machines.  If this is the case then adding additional computational requirements can be acceptable provided 
the communication costs are reduced through this addition of extra computation.  We have evaluated the performance impact 
of MDMP, and the communication verses computation trade-off on current HPC architectures where MDMP becomes beneficial, through
 benchmarking of our software, described in the next section.  However, we are still working on minimising the memory requirements 
for MDMP, as this will be important to ensure MDMP is usage on current and future HPC systems.  Furthermore, we should remember
that MDMP can be used as a simpler programming alternative to MPI with the optimised message passing functionality turned of at 
compile time, thereby removing all of these overheads if they are not beneficial for a given application of HPC platform.

\section{Experimental Setup}
\label{sec:5}

We evaluated the performance of the MDMP compared to standard C and MPI codes.  We undertook our evaluation using a range of common 
large scale HPC platforms and a set of simple, {\it kernel} style, benchmarks.  We have only evaluated the functionality using 2 nodes on each system, 
primarily testing the communications between a pair of communicating processors, one on each node.

\subsection{Computing Resources}

We used three different large scale HPC machines to benchmark performance.  The first was a {\bf Cray XE 6}, HECToR, is the UK National Supercomputing Service consists of 2816 nodes, each containing two 
16-core 2.3 GHz {\it Interlagos} AMD Opteron processors per node, giving a total of 32 cores per node, with 1 GB of memory per core. 
This configuration provides a machine with 90,112 cores in total, 90TB of main memory, and a peak performance of over 800 TFlop/s.  We used the PGI FORTRAN compile on HECToR, compiled with the \verb|-fastsse| 
optimisation flag.

The seconds was a {\bf Bullx B510}, HELIOS, which is based on Intel Xeon processors.  A node contains 2 Intel Xeon E5-2680 2.7 GHz processors giving 16-cores and 64 GB memory.  HELIOS 
is composed of 4410 nodes, providing a total of 70,560 cores and a peak performance of over 1.2 PFlop/s.  The network is built using Infiniband QDR non-blocking 
technology and is arranged using a fat-tree topology.  We used the Intel FORTRAN compiler on HELIOS, compiling with the \verb|-O2| optimisation flag.

The final resource was a {\bf BlueGene/Q}, JUQUEEN at Forsch\-ungszentrum Juelich.  JUQUEEN is a IBM BlueGene/Q system based on the IBM POWER architecture.
There are 28 racks composed of 28,672 nodes giving a total of 458,752 compute cores and a peak performance of 5.9 PFlop/s. 
Each node has an IBM PowerPC A2 processor running at 1.6 GHz and containing 16 SMT cores, each capable of running 4 threads, 
and 16 GB of SDRAM-DDR3 memory.   IBM's FORTRAN compile, xlf90, was used on JUQUEEN, compiling using the \verb|-O2| optimisation flag.

\section{Performance Results}
\label{sec:6}

We have been evaluating MDMP functionality using a number of different benchmarks.  Initially we tested 
the performance impact of instrumenting data reads and writes on a non-communicating code, the STREAMS
\cite{streams} benchmark, with the results presented in the first subsection below.  After this we evaluated 
the communications performance of MDMP verses communication implemented directly with MPI, the results of 
these evaluations are in the second subsection below.

Each of the STREAMS benchmark tests were repeated 10 times and an average runtime calculated.  For the communications benchmarks each operation was run 100 times, 
and each benchmark as repeated 3 times with the average time taken.

\subsection{Reference Implementation}

Whilst we have, in previous sections in this paper, outlined the principles of MDMP and how it designed to 
work, we do not yet have a full compiler based implementation of this functionality.  We have designed and 
implemented the runtime functionality that any compiler would added to a code when encountering MDMP 
pragmas, but have not yet implemented the compiler functionality.  Therefore, for this performance evaluation 
we are using benchmarks where the MDMP functionality has be implemented directly in the benchmark.

We have implemented two versions of MDMP; the first version implements all the required functionality within 
function calls to the MDMP library.  This includes data stores and lookups for all the data marked as 
being communicated within an MDMP communicating region. 

The second version implements exactly the same functionality but uses pre-processor macros to insert the 
required code directly into the source code, thereby removing the need for function calls at every point 
MDMP is used.  In the benchmark results this is named {\it Optimised MDMP}

Work is currently ongoing to implement a compiler based solution, utilising the LLVM\cite{llvm} compiler
infrastructure, to enable us to target all the main HPC computer languages with a single, full reference implementation.

\subsection{STREAM Benchmark Results}

STREAMS is often used to evaluate the memory bandwidth of computer hardware, and 
therefore was chosen as it will highlight any impact on the memory access and update efficiencies of 
computations when MDMP is added to a code.

The performance of the STREAMS benchmark was evaluated on all three of the hardware platforms we had available 
to us, although we are only presenting the results in the following tables from the Cray XE6 because, 
although the performance of the benchmark varies between machines, the relative performance difference between 
the original implementation of STREAMS and our MDMP implementations does not change significantly between these 
platforms.

We are reporting the results from a single process running on an otherwise empty node on the Cray XE6.
Table \ref{tab:mdmpstreambenchcommregion} (where Int stands for Integer and Db stands for Double) 
outlines the performance of the two versions of MDMP verses the 
original STREAMS code when the fully communicating functionality of MDMP is enabled and we have forced 
the MDMP library to treat each variable in the arrays being processed as if they were being communicated 
(i.e. we are fully tracking all the reads and writes to these array entries even though no communications 
are occurring).

\begin{savenotes}
\begin{table}[t]
\small{
\centering
\caption{C STREAM benchmark within a MDMP inside a communicating region (times in seconds)}
\begin{tabular}{|l|c|c|c|}\hline
Operation & Original  & MDMP & Optimised MDMP \\ \hline \hline
Int Assign &  0.000008 & 0.000172  & 0.000076\\ \hline 
Db Assign & 0.000010 & 0.000163 & 0.000084 \\ \hline 
Db Copy & 0.000009 & 0.000295 & 0.000151 \\ \hline 
Db Scale & 0.000016 & 0.000296 & 0.000152 \\ \hline 
Db Add & 0.000031 & 0.000447 & 0.000228  \\ \hline 
Db Triad & 0.000042 & 0.000439 & 0.00238 \\ \hline 
\end{tabular}
}
\label{tab:mdmpstreambenchcommregion}
\end{table}
\end{savenotes}

We can see from Table \ref{tab:mdmpstreambenchcommregion} that the MDMP functionality does have a 
significant impact on the overall runtime of all the benchmarks, adding around an order of magnitude increase 
to the runtime of the benchmark.  We would expect any benchmark like this, where no communications are 
involved and therefore there is no optimisation for MDMP to perform, to be detrimentally impacted by the 
additional functionality added in the MDMP implementation.  However, we can see that the optimised implementation of 
MDMP does not have as significant an impact as the original MDMP implementation.  As this is the first optimisation 
we have done to the MDMP functionality we are hopeful there is further scope for optimising the performance of MDMP and 
reducing the computational impact of the MDMP functionality.  

Furthermore, it is worth re-iterating that in this benchmark 
we are forcing MDMP to mark and track all the data in the STREAMS benchmark as if it is to be communicated.  MDMP is not designed 
to be beneficial for scenarios such as this, it is primarily designed to be useful in scenarios where a small amount of data 
(compared to the overall amount computed upon) is sent each iteration.  In a scenario such as the one this benchmark mimics 
(where all the data used in the computation is also communicated) MDMP should simply compiled down to the basic MPI calls as they 
would be more efficient in this scenario.  MDMP is designed to enable users to try different communication strategies, such 
as simply using the plain MPI calls, or trying to very the amount communication and computation intermingling, which enables 
users to experiment and evaluate which will give them the best performance for their application and use case.  Indeed, such functionality 
could also be built into the runtime of MDMP, enabling auto-tuning of the choice of communication optimisation on the fly.

We also ran the same benchmark with the code marked as outside a communicating region.  In this scenario, whilst the MDMP functionality has 
be enabled for all the variables in the calculation, the absence of a communicating region disables, at runtime, any data tracking 
associated with the variables.  Table \ref{tab:mdmpstreambenchnocommregion} presents the results for this benchmark.  We can see that 
the cost of the MDMP functionality has been substantially reduce, and indeed if we used the optimised functionality where the MDMP 
function calls have been removed and replaced with pre-processed code the MDMP performance is extremely close to the plain 
benchmark codes' performance.  This confirms that the MDMP functionality can be constructed in such a way as not to have a significant 
adverse impact on the key computational kernels of a code outside the places that communications are occurring.

\begin{savenotes}
\begin{table}[t]
\small{
\centering
\caption{C STREAM benchmark within a MDMP outside communicating region (times in seconds)}
\begin{tabular}{|l|c|c|c|}\hline
Operation & Original & MDMP & Optimised MDMP \\ \hline \hline
Int Assign &  0.000003 & 0.000103  & 0.000017 \\ \hline 
Db Assign & 0.000007 & 0.000079 & 0.000021 \\ \hline 
Db Copy & 0.000009 & 0.000118 & 0.000025 \\ \hline 
Db Scale & 0.000016 & 0.000126 & 0.000020 \\ \hline 
Db Add & 0.000030 & 0.000188 & 0.000025  \\ \hline 
Db Triad & 0.000038 & 0.000194 & 0.000025 \\ \hline 
\end{tabular}
}
\label{tab:mdmpstreambenchnocommregion}
\end{table}
\end{savenotes}

However, we can see from the results that if communications are present there is a significant performance impact on the data that is tracked by 
the MDMP functionality.  Our assumption is that the computational cost associated by MDMP can be more than offset by the reduction in communication 
costs for a program, but clearly this is dependent on the ratio between communications and computations for a given kernel, and the ratio of 
relative costs (in terms of overall runtime) of a communication verses a computation.

We evaluate the performance impact verses the communication cost savings in the next subsection, where we analyse some communication benchmarks.

\subsection{Message Passing Results}

We have constructed four simple benchmarks to evaluate MDMP against MPI.  The first is a {\bf PingPong} benchmark where a process sends a message to 
another process who copies the received data from the receive buffer into it's send buffer and sends it back to the first process, who performs the same copying process and sends it back again.  
This pattern is repeated many times and the time for the communications are recorded.  The benchmark can send a range of message sizes.  For the reference MPI benchmark only a single message is 
sent each iteration of the benchmark containing the fully amount of data to be sent.  For the MDMP version the \verb|send| and \verb|recv| functionality specifies the single message to be 
sent and received, and performs the send and receive on the first iteration of the benchmark but on subsequent iterations of the benchmark the MDMP functionality identifies when each element of the 
message data is ready to be sent (through tracking the data copying process between the send and receive buffers) and sends individual elements when they are ready to go.  This will mean that 
for a run of the benchmark using a message of 1000 elements in size the MPI version will send one message between processes whereas the MDMP version will send 1000 message (apart from on the 
first iteration where it will only send one message).

The second benchmark, called {\bf SelectivePingPong} alters the basic PingPong benchmark we have already described by performing the same functionality but only sending a portion of the 
overall data owned by a process in the messages.  It is possible to vary both the overall size of data each process has, and the amount of that data that is 
sent, for instance you could have each process having an array that is 100 elements long but only the first 10 and last 10 elements are sent in the PingPong 
messages.  This benchmark is designed to investigate the performance impact of varying the overall data in a computation and the amount that is being communicated 
via MDMP.

The third benchmark, called {\bf DelayPingPong}, also alters the basic PingPong benchmark by adding a delay in the loop that copies the data from the receive buffer to the send buffer.  This delay is variable and is 
designed to simulate some level of computational work being undertaken during what would be the main computational loop for a computational kernel using MDMP.  The delay is performed by a routine which iterates 
through a loop adding an integer to a double a specified number of times (delay elements).

The final benchmark, {\bf SelectiveDelayPingPong}, combines the second and third benchmarks meaning the PingPong process can contain both user defined delay in the data copy loop and a selective amount of 
data to be transferred.

\begin{figure*}[t]
\centering
\subfloat[PingPong Benchmark]{\epsfig{file=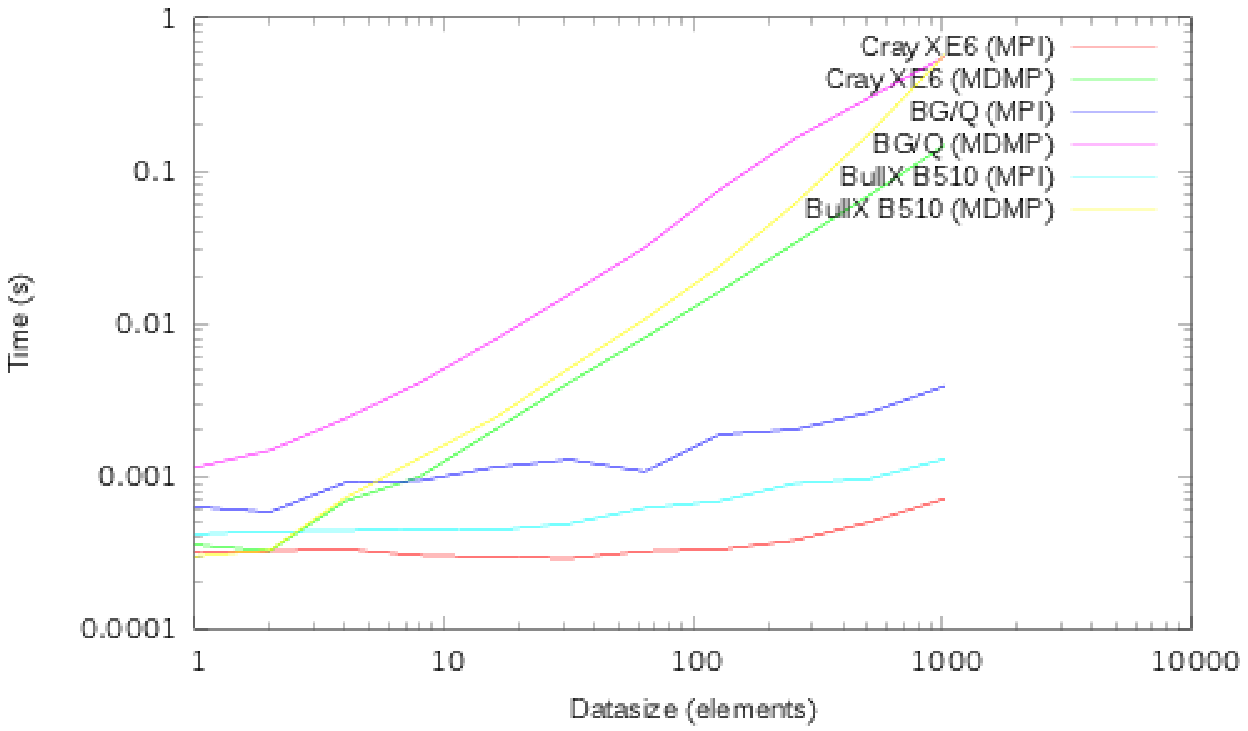,width=0.5\textwidth}\label{fig:combinedpingpong}}
\subfloat[DelayPingPong Benchmark (1024 elements)]{\epsfig{file=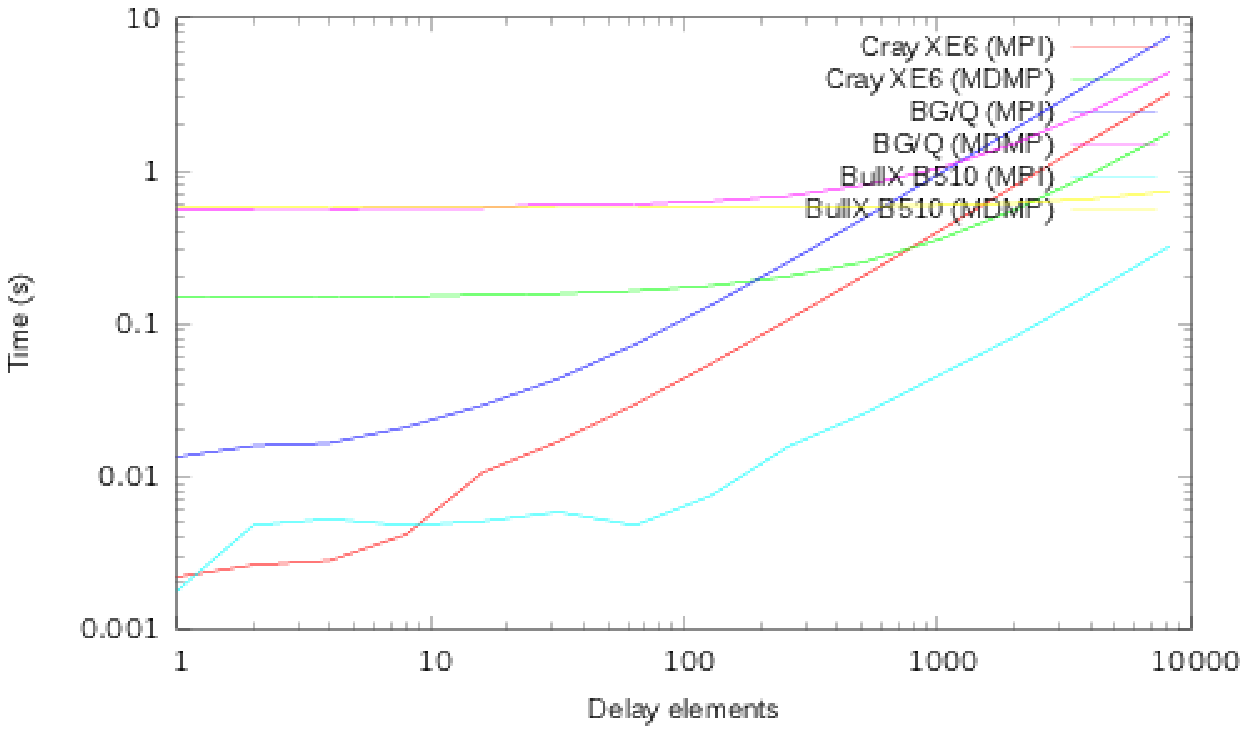,width=0.5\textwidth}\label{fig:combinedpingpongdelay}}
\caption{Runtime of PingPong and DelayPingPong benchmarks using up to 1024 data elements}
\end{figure*}

Figure \ref{fig:combinedpingpong} demonstrates the cost of MDMP compared to plain MPI where there is no scope for communication and computational overlaps.  The runtime for MDMP increases more or less linearly as the 
size of the data to be transferred increases, whereas the runtime for MPI stays relatively constant.

However, if we examine Figure \ref{fig:combinedpingpongdelay} we can see that MDMP begins to see some benefits over MPI when the delay added to the data 
copy routine is increased.  The JUQUEEN and HECToR MDMP is faster than MPI when the delay elements are around 1000 and 800 elements respectively, although 
for HELIOS MPI is always faster than MDMP (albeit with a smaller gap in performance between the two methods).

\begin{figure*}[t]
\centering
\subfloat[SelectivePingPong]{\epsfig{file=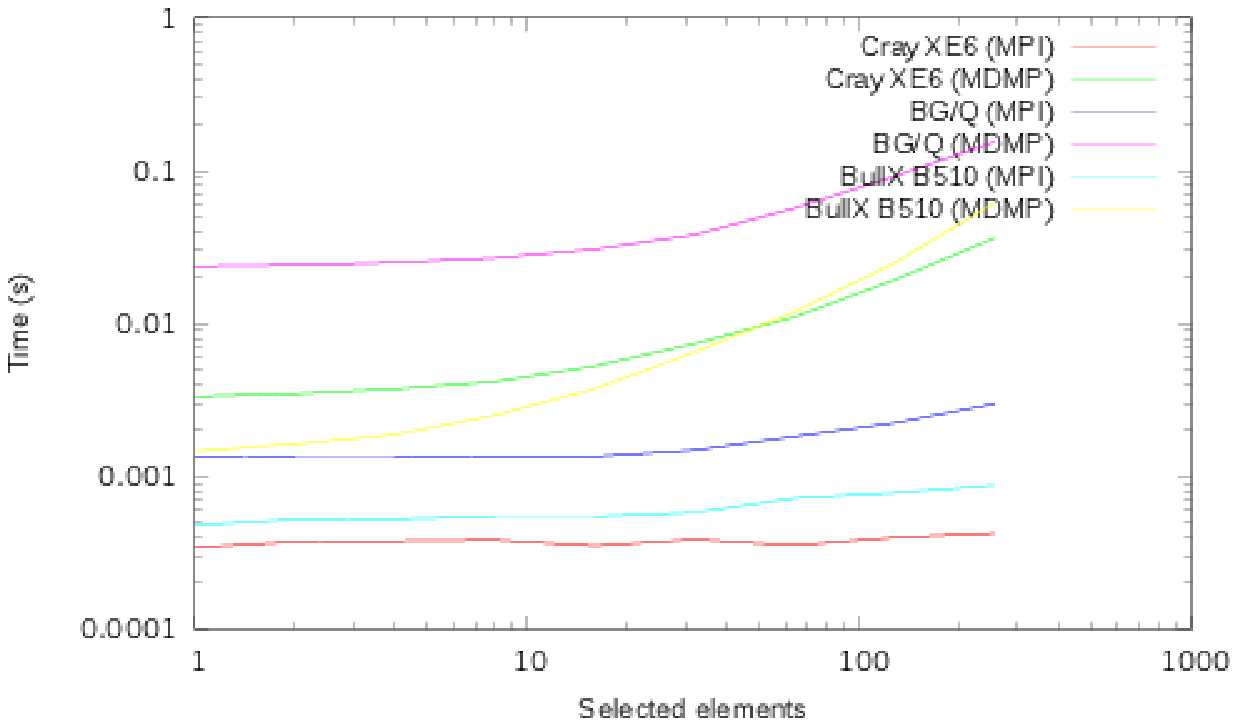,width=0.5\textwidth}\label{fig:combinedpingpongselective1024}}
\subfloat[SelectiveDelayPingPong]{\epsfig{file=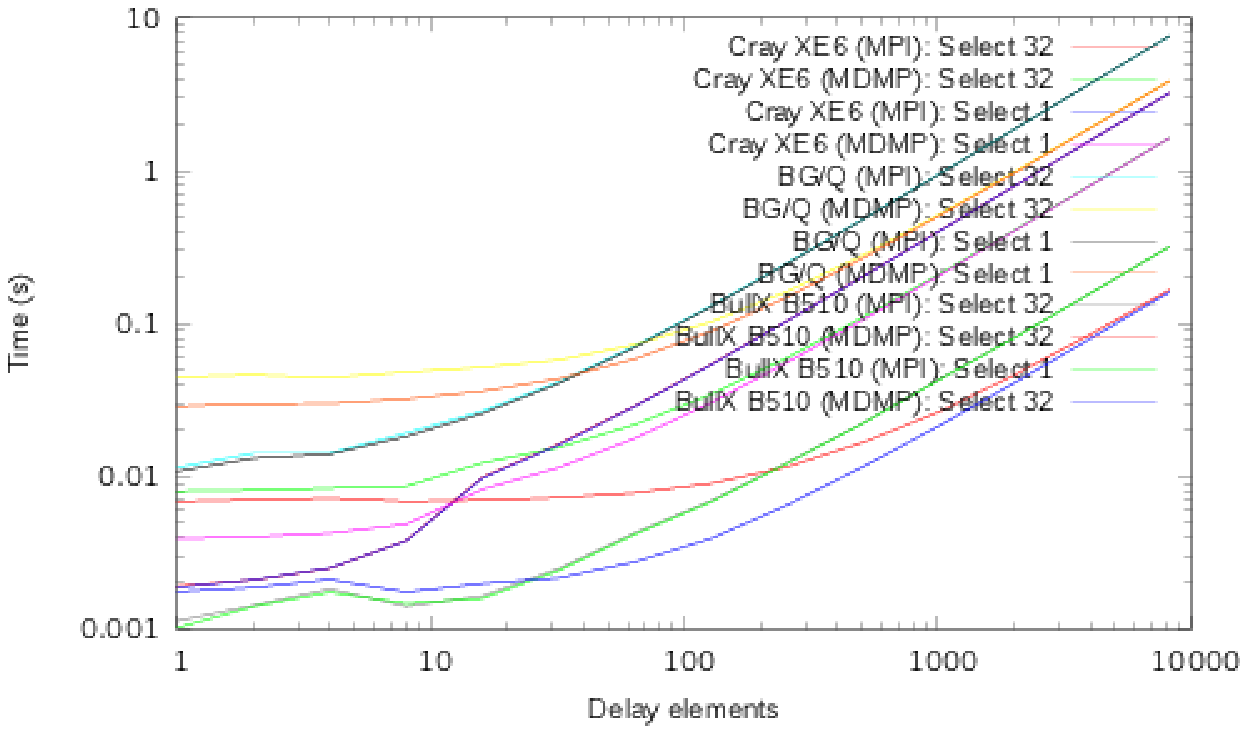,width=0.5\textwidth}\label{fig:combinedpingpongselectivedelay1024}}
\caption{Runtime of benchmarks with 1024 data elements, varying the number of selected elements or the number of delay elements}
\end{figure*}

If not all the data that is copied between buffers is sent, as in the case of the SelectivePingPong benchmark shown in Figure \ref{fig:combinedpingpongselective1024}, 
then in comparison to the normal PingPong benchmark the overall difference in performance is reduced between MPI and MDMP although MDMP is still more costly than 
MPI.

Finally, the combined benchmark, results shown in \ref{fig:combinedpingpongselective1024} where 1024 overall data elements are processed and either 1 or 32 elements are 
sent with variable amounts of delays, highlight where MDMP can improve performance.  When only one element is being sent then all it requires is 16 floating point adds between 
communications (16 delay elements)\footnote{Actually there are $1023*16$ delay elements as there is a delay per array element} to enable MDMP to optimise communications.  If 32 
elements are being sent then around 32 floating point adds are required to enabling the communication hiding that MDMP enables to provide a performance benefit.

Whilst these benchmarks are beneficial in enabling us to evaluate MDMP performance we recognise that a more realistic benchmark that evaluates MDMP performance 
against real kernel computations would also be useful as it would enable us to evaluate the overall impact of MDMP on cache, memory, and processor usage for real 
applications.  We are in the process of undertaking such benchmarks at the moment but unfortunately do not have these results in time for this paper submission.

\section{Conclusions}

We have outlined a novel approach of message passing programming on distributed memory HPC architectures and demonstrated that, given a reasonable level of 
computations to the communications to be performed, MDMP can reduce the overall cost of communications and improve application performance.  We are aware the 
MDMP presents performance risks for parallel programs, including impacting cache and memory usage, and consuming additional memory.  However, we belief the ability to 
enable and disable MDMP optimisations, and the potential benefits to ease of use and programmability from MDMP, make this approach a sensible one to 
investigate future message passing programming.

We are currently working on a full compiler implementation of MDMP, including a formal MDMP language definition, and more involved benchmarks to evaluate MDMP 
in much more detail.

\section{Acknowledgments}
Part of this work was supported by an e-Science grant from Chalmers University.

Part of this work was supported by the Nu-FuSE project which 
is funded by EPSRC through G8 Multilateral Research Funding Nu-FuSE grant EP/J004839/1.

\balancecolumns
\end{document}